\documentclass[12pt,draftclsnofoot,onecolumn]{IEEEtran}

\usepackage{amssymb}
\usepackage{amsmath}
\usepackage{amsfonts}
\usepackage{graphicx}
\usepackage{subfigure}
\usepackage{cite}
\usepackage{enumerate}
\usepackage{url}
\usepackage[ruled]{algorithm2e}
\usepackage{algorithmicx}
\usepackage{mathtools}
\usepackage{cases}
\usepackage{subeqnarray}
\allowdisplaybreaks[4]

\begin{document}
\title{\LARGE{Non-orthogonal Multiple Access for High-reliable and Low-latency V2X Communications}}
\author{
\IEEEauthorblockN{
\normalsize{Boya Di}\IEEEauthorrefmark{1},
\normalsize{Lingyang Song}\IEEEauthorrefmark{1},
\normalsize{Yonghui Li}\IEEEauthorrefmark{2},
\normalsize{and Geoffrey Ye Li}\IEEEauthorrefmark{3}\\}
\IEEEauthorblockA{
\IEEEauthorrefmark{1}\normalsize{School of Electrical Engineering and Computer Science, Peking University, China.} \\
\IEEEauthorrefmark{2}School of Electrical and Information Engineering, The University of Sydney, Australia.\\
\IEEEauthorrefmark{3} School of Electrical and Computer Engineering, Georgia Institute of Technology, USA.}\\
}
\maketitle

\begin{abstract}
In this paper, we consider a dense vehicular communication network where each vehicle broadcasts its safety information to its neighborhood in each transmission period. Such applications require low latency and high reliability, and thus, we propose a non-orthogonal multiple access scheme to reduce the latency and to improve the packet reception probability. In the proposed scheme, the BS performs the semi-persistent scheduling to optimize the time scheduling and allocate frequency resources in a non-orthogonal manner while the vehicles autonomously perform distributed power control. We formulate the centralized scheduling and resource allocation problem as equivalent to a multi-dimensional stable roommate matching problem, in which the users and time/frequency resources are considered as disjoint sets of players to be matched with each other. We then develop a novel rotation matching algorithm, which converges to a $q$-exchange stable matching after a limited number of iterations. Simulation results show that the proposed scheme outperforms the traditional orthogonal multiple access scheme in terms of the latency and reliability.
\end{abstract}

\begin{IEEEkeywords}
Non-orthogonal multiple access, V2X broadcasting, resource allocation, scheduling problem, matching game.
\end{IEEEkeywords}

\section{Introduction}
With the rapid development of intelligent transportation systems (ITS), a growing number of vehicular applications have emerged to provide a safer and more efficient driving experience for our daily life. Among various applications, safety critical services play a vital role in the blueprint of the future ITS, supported by the vehicle-to-everything (V2X) communications~\cite{GOEGBKT-2011}, including the vehicle-to-vehicle (V2V), vehicle-to-pedestrian (V2P), and vehicle-to-network/infrustructure~(V2N/I). To achieve low-latency and high-reliability~(LLHR) for the V2X services, recently the widely deployed Long Term Evolution (LTE) networks have been considered as a very promising solution to achieve large cell coverage, controllable latency, and high data rates even in a high-mobility scenario~\cite{SS-2010}.

The LTE-based V2X services combine the ad hoc and cellular network architecture by exploiting the device-to-device (D2D) communications~\cite{LDZE-2015}. Hence, the end-to-end latency can be reduced compared to the cellular uplink/downlink (UL/DL) mode, and the quality of services can be guaranteed in contrast to the 802.11p~\cite{ACCIM-2015}. However, unlike the traditional D2D communications, V2X applications require stringently low latency, which poses new challenges to the LTE-based vehicular network, especially in a dense network causing severe data congestion~\cite{SHPPZF-2016}. One of the main reasons is that the existing LTE networks are based on the orthogonal multiple access (OMA), and the limited spectrum resource have not been fully and efficiently utilized~\cite{LYZH-2017}, leading to the severe data congestion and low access efficiency in a dense network. Therefore, a more spectrally efficient radio access technology is required for the V2X services.

To handle the challenges of access collisions and massive connectivity, non-orthogonal multiple access (NOMA) schemes have been introduced as a potential solution, which allow users to access the channel non-orthogonally by either power-domain~\cite{SKBNLH-2013} or code-domain multiplexing~\cite{huawei-wp}. Multiple users with different types of traffic requests can transmit concurrently on the same channel to improve spectrum efficiency and alleviate the congestion of data traffic, thereby reducing the latency. To make the NOMA scheme more practical, various multi-user detection (MUD) techniques, such as successive interference cancelation (SIC)~\cite{HBCJMKL-2014}, have been applied at the end-user receivers to cope with co-channel interference caused by spectrum sharing among various users.


Capable of achieving high overloading transmission over limited resources, NOMA provides a new dimension for V2X services to alleviate the traffic congestion, thereby reducing the latency. In this paper, we consider the V2X broadcast scenario~\cite{3GPP-2016} where every vehicle needs to broadcast the safety information to its neighborhood\footnote{V2X broadcasting refers to V2V/I broadcasting for safety information.} in each transmission period. Each period consists of multiple time slots and the transmitter-receiver (Tx-Rx) selection\footnote{A vehicle is called a Tx user if it broadcasts in a time slot and a Rx user in other time slots.} needs to be determined for each slot such that all the vehicles can update their safety information in at least one slot during this period. Sub-channel allocation in each time slot is performed to manage the co-channel interference caused by the non-orthogonal nature. Moreover, to perform joint decoding in an non-orthogonal manner, a different method of the real-time power control is applied by the users, in contrast to the traditional OMA-based case. New challenges are thus posed in the design of scheduling and resource allocation schemes. Meanwhile, the LLHR requirement and dense topology of vehicular networks need to be considered.

In this paper, we propose a NOMA-based mixed centralized/distributed (NOMA-MCD) scheme for the V2X broadcasting system. The centralized semi-persistent scheduling (SPS)~\cite{LM-2013} is performed by the BS where the Tx-Rx is selected and the time and frequency resources are allocated every few transmission periods. For the BS, Tx-Rx selection and resource allocation problem can be formulated as a non-linear integer programming problem to maximize the packet reception probability of the network. Afterwards, the autonomous distributed power control is performed by each user, and a user association scheme is designed, where the Tx users utilize the control signals to iteratively adjust their transmit power based on the feedback of the Rx users.

To tackle such a combinatorial optimization problem on Tx-Rx selection and resource allocation, we then decouple it into two multi-dimensional stable roommate (MD-SR) matching problems~\cite{DR-1989}, in which the vehicles and time slots/sub-channels are considered as two disjoint sets of ``students" and ``rooms" such that multiple ``students" can occupy the same ``room". It is not trivial to find an efficient algorithm to solve such a NP-hard MD-SR problem since most existing matching algorithms~\cite{DR-1989,7AM-1992,ESAKJV-2009,M-2013} are designed for the two-dimensional SR problem in which at most two ``students" can share one ``room". Therefore, we develop a novel rotation matching algorithm for the MD-SR problem which converges to a stable matching.

Only limited works have discussed how to improve the performance of the safety critical applications from a NOMA-based perspective. The most related ones~\cite{WPL-2016,WDEF-2016,CGMH-2014} focus on either the LTE-based D2D broadcasting systems or the cellular V2V unicast systems with the assumptions of fixed Tx-Rx selection or transmit power. In~\cite{WPL-2016}, a distributed scheme for the coordination of the D2D broadcast transmission has been presented in which the traffic collision has been avoided through the orthogonal use of the spectrum resources in the neighborhood of each Tx user. In~\cite{WDEF-2016}, the resource management for D2D unicast safety-critical vehicular communications has been discussed and the sum rate of cellular mobile users has been guaranteed under the constraints of satisfying the vehicles' requirements on latency and reliability. In~\cite{CGMH-2014}, a centralized sub-channel allocation scheme involving several D2D broadcast groups underlaying cellular network has been proposed by utilizing a greedy algorithm.

To sum up, the main contributions of our work are listed below:
\begin{itemize}
\item We propose a novel NOMA-MCD scheme for the V2X broadcasting system combining the centralized SPS at the BS and the distributed power control of the vehicles. The UL latency can be reduced and the reliability can be further improved.
\item We consider the centralized Tx-Rx selection and the resource allocation problem as a packet reception probability maximization problem, and decouple it into two MD-SR matching problems, which can be solved by a novel rotation matching algorithm.
\item An iterative user association scheme is proposed to solve the distributed power control problem of the Tx users, in which the Tx and Rx users exchange control messages for joint decoding and transmit power adjustment.
\end{itemize}

The rest of this paper is organized as follows. In Section \uppercase\expandafter{\romannumeral2}, we describe the system model of the NOMA-based cellular V2X broadcasting system. In Section \uppercase\expandafter{\romannumeral3}, we elaborate the proposed NOMA-MCD scheme, and formulate the centralized scheduling and resource allocation problem of the BS as a packet reception probability maximization problem. An iterative user association scheme is designed for the distributed power control problem of the Tx users. In Section \uppercase\expandafter{\romannumeral4}, we reformulate and decouple the maximization problem into two MD-SR problems and then we propose a rotation matching algorithm to solve them. In Section \uppercase\expandafter{\romannumeral5}, the system performance is analyzed and the properties of the algorithms are investigated. Simulation results are presented in Section \uppercase\expandafter{\romannumeral6}, and finally, we conclude the paper in Section \uppercase\expandafter{\romannumeral7}.



\section{System Model}%
In this section, we present the system model of the cellular V2X broadcasting system and propose the NOMA-based cellular V2X broadcasting scheme to reduce the access collision.
\subsection{Scenario Description}

\begin{figure}[!t]
\centering
\includegraphics[width=5in]{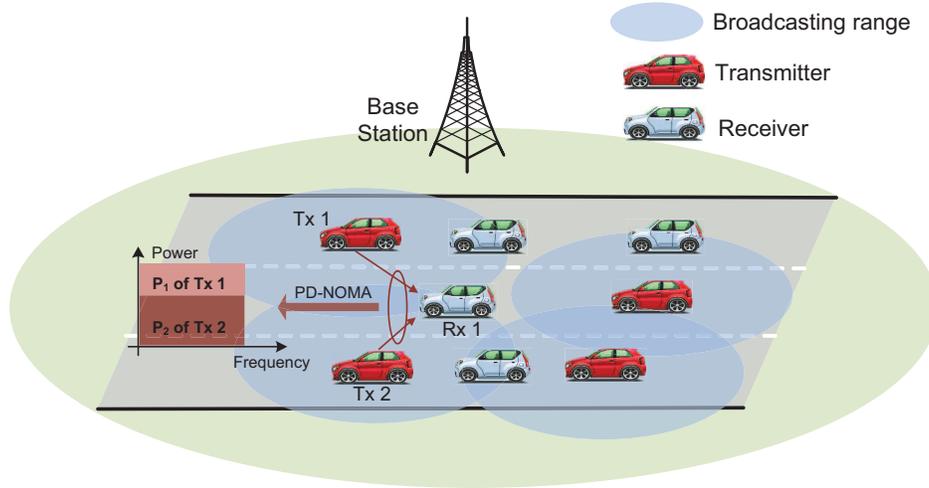}
\caption{System model of V2X transmission.} \label{system_model}
\end{figure}

Consider an urban V2X broadcast system as shown in Fig.~\ref{system_model}. In every short transmission period consisting of multiple time slots, each of $N$ vehicles broadcasts safety-critical information\footnote{For example, the cooperative awareness messages record periodic time-triggered position information with transmission interval of $100$ms.} to its neighborhood in at least one time slot. The pedestrians always act as Rx users in each time slot. Direct data transmission between neighboring users is achieved in the D2D mode, i.e., users in proximity can communicate with each other directly bypassing the BS. The available bandwidth is divided into $K$ sub-channels for transmitting.

Due to the dense topology of the network, when more than one Tx user (e.g., Tx user 1 and Tx user 2 in Fig.~\ref{system_model}) are assigned the same time-frequency resources, severe collision may occur for those Rx users (e.g., Rx user 1) locating in the overlapping region of two adjacent Tx users' communication ranges. To reduce the probability of the collision as well as the transmission delay caused by it, non-orthogonal allocation of radio resources is then considered such that one sub-channel can be occupied by multiple Tx users simultaneously~\cite{BLY-2016}. Each conflicting Rx user (such as Rx user 1 in Fig.~\ref{system_model} mentioned above) utilizes the SIC technique to decode the received superposed signals, thereby avoiding the collision.

\subsection{NOMA-based Collision Avoidance}
\subsubsection{Channel Model}
Based on the NOMA scheme, the received signal of Rx user $m$ over subchannel $k$ in time slot $i$ (the $i$th time slot of a transmission period) can then be presented by
\begin{equation} \label{receiving_signal}
y_{m,k}^{(i)} = \sum\limits_{j \in {\mathcal{N}_m^{(i)}}} {\gamma _{j,k}^{(i)}\sqrt {p_{j,k}^{\left( i \right)}}{H_{j,m,k}^{(i)}}{s_j^{(i)}}}  + {n_m^{(i)}},
\end{equation}
in which ${\mathcal{N}_m^{(i)}} = \left\{ {1 \leq j \leq N|{d_{j,m}^{(i)}} \le {r}} \right\}$ denotes the set of users within Rx user $m$'s communication range of interest, i.e., a disk with the radius of $r$, $\gamma_{j,k}^{(i)}$ is a binary variable to indicate whether user $j$ is a Tx user transmitting over subchannel $k$ in time slot $i$, $p_{j,k}^{(i)}$ is the transmitted power of Tx user $j$, $H_{_{j,m,k}}^{(i)}$ denotes the channel coefficient of subchannel $k$ between Tx user $j$ and Rx user $m$ in time slot $i$, $s_j^{(i)}$ represents the transmitted symbol of Tx user $j$ in time slot $i$, ${n_m^{(i)}} \sim \mathcal{CN}\left( {0,{\sigma _n}^2} \right)$ is the additive white Gaussian noise~(AWGN) for Rx user $m$, and ${\sigma _n}^2$ is the noise variance.

In equation $\left( \ref{receiving_signal} \right)$, the channel coefficient $H_{_{j,m,k}}^{(i)}$ can be defined as $H_{_{j,m,k}}^{(i)} = h_{j,m,k}^{(i)}{g_{j,m}^{(i)}}$, in which $h_{j,m,k}^{(i)}$ and $g_{j,m}^{(i)} = \beta {\left( {d_{j,m}^{(i)}} \right)^{ - \alpha }}$ are the Rayleigh fading and distance-dependent pathloss coefficient, respectively, with $d_{j,m}^{(i)}$ denoting the distance between two users $j$ and $m$ in time slot $i$. By utilizing the periodic velocity and location information updated by the users, $d_{j,m}^{(i)}$ can be predicted by the BS,
\begin{equation}
d_{j,m}^{(i)} = i\cdot\frac{{\left({\textbf{\emph{v}}_{j,m}^{(0)}}\right)^T \left( {{{\textbf{\emph{x}}}_m^{(0)}} - {{\textbf{\emph{x}}}_j^{(0)}}} \right)}}{{{{\left\| {{{\textbf{\emph{x}}}_m^{(0)}} - {{\textbf{\emph{x}}}_j^{(0)}}} \right\|}_2}}}, i \ge 1,
\end{equation}
in which $\emph{\textbf{v}}_{j,m}^{(0)}$ is the relative velocity vector between user $j$ and $m$ at the beginning of a transmission period, i.e., in time slot 0, and $\textbf{\emph{x}}_j^{(0)}, \emph{\textbf{x}}_m^{(0)} \in \mathbb{R}^2$ denotes the positions of user $j$ and $m$ in time slot 0.

\subsubsection{SIC Decoding}
According to NOMA, each conflicting Rx user $m$ decodes the received signals in a decreasing order of channel gains~\cite{DP-2005}. For example, consider two Tx users $j$ and $j'$ occupying subchannel $k$ in time slot $i$ transmitting to Rx user $m$ satisfying ${\left| {H_{j,m,k}^{(i)}} \right|^2} < {\left| {H_{j',m,k}^{(i)}} \right|^2}$. The conflicting Rx user $m$ first detects the data signals of Tx user $j'$. After subtracting the decoded signals, it then decodes the signal from Tx user $j$ while regarding the signals of those Tx users with lower channel gains than that of Tx user $j$ as noise. For any conflicting Rx user $m$, the achievable rate obtained from Tx user $j$ over subchannel $k$ in time slot $i$ can then be formally presented~\cite{SKBNLH-2013,BLY-2016}:
\begin{equation} \label{rate_equation}
R_{_{j,m,k}}^{(i)} = {\log _2}\left( {1 + \frac{{{p_j^{(i)}}\rho _{_{j,m,k}}^{(i)}}}{{1 + \sum\limits_{j' \in \mathcal{S}_{_{j,m,k}}^{(i)}} {{p_{j'}^{(i)}}\rho _{_{j',m,k}}^{(i)}} }}} \right),
\end{equation}
where ${\rho _{j,m,k}^{(i)}} = {\left| {{H_{j,m,k}^{(i)}}} \right|^2}/\left({n_m^{(i)}}\right)^2$ represents the SNR of the Tx user $j$ -- Rx user $m$ link, and $\mathcal{S}_{_{j,m,k}}^{(i)}$ is the set of active Tx users causing interference when Rx user $m$ decodes the signal of Tx user $j$.

\subsubsection{Criteria for Successful Decoding}
We assume that the rate threshold for a Rx user to decode the signal of a Tx user is ${\bar R}_{th}$. Therefore, when any conflicting Rx user $m$ receives from multiple Tx users, two criteria for Rx user $m$ to successfully detect the signal $\emph{\textbf{x}}_j$ received from Tx user $j$ over subchannel $k$ in time slot $i$ are:
\begin{itemize}
\item The signals of other Tx users with higher channel gains than Tx user $j$ are successfully decoded first;
\item The rate threshold of Tx user $j$ is satisfied, i.e., $R_{_{j,m,k}}^{(i)} \geq {{\bar R}_{th}}$.
\end{itemize}
The above criteria can be mathematically expressed as
\begin{equation} \label{decode_success}
\prod\limits_{j' \in \mathcal{N}_m^{(i)}\backslash \mathcal{S}_{j,m,k}^{(i)}} {\gamma _{j,k}^{(i)}{{\left( {R_{j',m,k}^{(i)} - {{\bar R}_{th}}} \right)}^ + }}  = 1,
\end{equation}
in which $\left( \cdot \right)^+$ is a signed function and $j' \in \mathcal{N}_m^{(i)}\backslash \mathcal{S}_{j,m,k}^{(i)}$ implies that the channel gain between an active Tx user ${j'}$ and Rx user $m$ is higher than that between Tx user $j$ and Rx user $m$.

\subsection{NOMA-based Mixed Centralized/Distributed Scheme}
Based on the above non-orthogonal manner of reducing the collision, we aim to design a scheme in which each user can successfully broadcast the safety information to as many neighboring users as possible while satisfying the latency requirement. According to equations $\left( \ref{rate_equation} \right)$ and $\left( \ref{decode_success} \right)$, system performance relies on Tx-Rx selection and frequency/power-domain resource allocation. Therefore, the following key problems need to be solved in the NOMA-based scheme design:
\begin{itemize}
\item How to determine the role of each user (i.e., Tx/Rx user) in each time slot of a transmission period;
\item How to allocate the subchannels to the set of Tx users;
\item How to perform the power control of the Tx users in each time slot.
\end{itemize}
To address these issues, we propose a NOMA-MCD scheme consisting of the following two phases: 1) centralized Tx-Rx selection and time-frequency resource allocation of the BS; 2) distributed power control of the users.

\subsubsection{Centralized Tx-Rx Selection and Time-frequency Resource Allocation of the BS}
In the traditional dynamic scheduling, resource allocation may cause significantly large delay since the users need to send resource request messages to the BS for every data packet. To reduce such UL latency, centralized SPS at the BS is considered in which a predefined sequence of resources are allocated to the users by the BS at the beginning of each SPS period consisting of one or more transmission periods. The resource allocation scheme keeps unchanged during each transmission period of the SPS period. The SPS is considered as a suitable scheme for periodic short message transmission with a fixed beacon rate~\cite{LM-2013}.

To take full advantage of the users' global position information obtained by the BS from the periodic user broadcasting, the BS determines the Tx-Rx selection and subchannel allocation of each time slot at the beginning of each SPS period in the centralized NOMA-based SPS scheme. It also provides a more stable performance of latency and reliability compared to a distributed scheme in which the users compete with each other to access the channel.

\subsubsection{Distributed Power Control of the Users}
To benefit from the power domain multiplexing, the NOMA scheme requires prior knowledge of the users for SIC decoding such as the real-time CSI, which is hard for the BS to obtain due to the mobility of the vehicles. Distributed real-time power control of the users is then performed to achieve a better decoding effect after the centralized SPS at the BS. In each time slot, the transmitted power of the Tx users is adjusted based on feedback sent from the conflicting Rx users via the control signaling.

Design of the above two phases will be presented in detail in Section \uppercase\expandafter{\romannumeral3} and \uppercase\expandafter{\romannumeral4}, respectively.



\section{NOMA-based Centralized Semi-Persistent Scheduling}
In this section, we formulate the centralized Tx-Rx selection and time-frequency resource allocation of the BS as a packet reception probability (PRP) maximization problem and then solve it by utilizing the matching theory.

\subsection{Problem Formulation}
Note that in the vehicular network, not only the full CSI is very costly for the BS to acquire, but also CSI can become easily outdated due to the rapid variation of the small scale fading caused by the mobility of vehicles~\cite{CYS-2015}. Therefore, only partial CSI containing the path loss and shadowing\footnote{Note that the assumption of partial CSI is widely used in vehicular networks such as~\cite{CYS-2015,WDEF-2016}.}, i.e., $H_{_{j,m,k}}^{(i)} = {g_{j,m}^{(i)}}$, can be used for the centralized resource allocation.

To improve the reliability of the network, we try to alleviate the collision such that fewer retransmissions are required, which in turn reduces the latency. Since the collision reduction can improve the PRP of the network, we then aim at maximizing the PRP of each transmission period, which is proportional to the total number of signals successfully decoded by all Rx users.

Whether Rx user $m$ can successfully decode the signal of Tx user $j$ can be judged according to $\left( \ref{decode_success} \right)$. The number of successfully decoded signals within one transmission period can then be given by
\begin{equation}
\sum\limits_{i = 1}^{{T_v}} {\sum\limits_{k = 1}^K {\sum\limits_{m = 1}^N {\left( {1 - \gamma _{m,k}^{\left( i \right)}} \right)\sum\limits_{j \in {\cal{N}}_m^{\left( i \right)}} {\gamma _{j,k}^{\left( i \right)}\prod\nolimits_{j' \in {\cal N}_m^{\left( i \right)}\backslash {\cal S}_{j,m,k}^{\left( i \right)}} {\gamma _{j',k}^{\left( i \right)}{{\left( {R_{j',m,k}^{\left( i \right)} - {{\bar R}_{th}}} \right)}^ + }} } } } },
\end{equation}
in which $T_v$ is the total number of time slots in each transmission period. However, we note that the decoding effect of SIC is closely related to the gap between the signal strength of multiple Tx users in practice. In other words, a Rx user can decode two Tx users with a larger gap of signal strength more easily due to the SIC principle~\cite{ZPH-2016}, which cannot be accurately reflected by the binary function. To be more general, we adopt the logistic function instead of the indicator function to better describe the probability that the signal of Tx user $j$ is successfully decoded by Rx user $m$ over subchannel $k$ in time slot $i$, given by
\begin{equation} \label{utility}
U_{j,m,k}^{\left( i \right)} = \gamma _{j,k}^{\left( i \right)}\left( {1 - \gamma _{m,k}^{\left( i \right)}} \right){\prod _{j' \in {\cal N}_m^{\left( i \right)}\backslash {\cal S}_{j,m,k}^{\left( i \right)}}}\frac{{\gamma _{j',k}^{\left( i \right)}}}{{1 + {e^{ - \eta \left[ {R_{j',m,k}^{\left( i \right)} - {{\bar R}_{th}}} \right]}}}},
\end{equation}
where $\eta$ is the slope parameter of the logistic function. The Tx-Rx selection and time-frequency resource allocation problem of the BS in one transmission period can then be formulated as below:
\begin{align}
&\mathop {\max }\limits_{\left\{ {\gamma _{j,k}^{\left( i \right)}} \right\}} \sum\limits_{i = 1}^{{T_v}} {\sum\limits_{k = 1}^K {\sum\limits_{m = 1}^N {\sum\limits_{j \in {\cal{N}}_m^{\left( i \right)}} {U_{j,m,k}^{\left( i \right)}} } } } \label{system_optimization}\\
s.t.~  &\sum\limits_{k = 1}^K {\left( {\gamma _{j,k}^{\left( i \right)} + \gamma _{j',k}^{\left( i \right)}} \right)}  \le 1,i = 1,...,{T_v},\left\{ {j,j'} \right\} \in \left\{ {1 \le j,j' \le N|d_{j,j'}^{\left( i \right)} < r} \right\},\tag{\ref{system_optimization}$a$}\\
&\sum\limits_{k = 1}^K {\gamma _{j,k}^{(i)}}  \le {K_{\max }},i = 1,\cdots,{T_v},j = 1,\cdots,N,\tag{\ref{system_optimization}$b$}\\
&1 \le \sum\limits_{i = 1}^{{T_v}} {\sum\limits_{k = 1}^K {\gamma _{j,k}^{\left( i \right)}}}  \le {T_{\max }}, j = 1, \cdots, N,\tag{\ref{system_optimization}$c$}\\
&{\sum _{j \in {\cal N}_m^{\left( i \right)}}}\gamma _{j,k}^{(i)}\left( {1 - \gamma _{m,k}^{\left( i \right)}} \right) \le {K_u},m = 1, \cdots ,N,k = 1, \cdots ,K,i = 1,...,{T_v}, \tag{\ref{system_optimization}$d$}\\
&\gamma _{j,k}^{(i)} = \left\{ {0,1} \right\}, j = 1, \cdots ,N, k = 1, \cdots ,K, i = 1,\cdots,{T_v}.\tag{\ref{system_optimization}$e$}
\end{align}
Constraint $\left( 7a \right)$ denotes that any two users within each other's communication range cannot be Tx users at the same time since they will never receive each other's message while transmitting due to the half duplex nature. For the sake of user fairness, a Tx user can occupy no more than $K_{max}$ subchannels within one time slot and at most $T_{max}$ time slots are allocated to it in one transmission period as expressed in constraints $\left( 7b \right)$ and $\left( 7c \right)$. Considering the decoding complexity of SIC at the receiver, we assume that each subchannel can be assigned to at most $K_u$ Tx users with overlapping communication disks simultaneously as presented in constraint $\left( 7d \right)$.


\subsection{Matching Algorithm Design}

As discussed in Appendix A, the formulated problem in $\left( \ref{system_optimization} \right)$ is non-convex and NP-hard. To solve this problem, we design a novel multi-dimensional (MD) rotation matching algorithm. We first start with the traditional two-dimensional stable roommate (SR) problem~\cite{GSB-2015} in which 2$n$ students aim at selecting roommates given $n$ identical rooms~\cite{DR-1989,7AM-1992}.

Based on the above SR problem, we consider the set of users and time slots/subchannels as two disjoint sets of ``students" and ``rooms". Note that one time slot cannot be assigned to two neighbouring Tx users due to the half-duplex nature as expressed in constraint $\left( 7a \right)$. Therefore, time scheduling is considered prior to sub-channel allocation for the users and the time-frequency resources cannot be considered as a whole. Considering the above relation between the time slots and frequency resources, we formulate the optimization problem in $\left( \ref{system_optimization} \right)$ as two MD versions of the SR problem. Different from the 2D version, one ``room" can be occupied by multiple ``students" and the complicated interaction between multiple ``students" is difficult to be captured via the traditional matching algorithms, thereby bringing new challenge to the problem.

To address the above challenge, we propose a rotation matching algorithm for the MD user-time matching problem and then extend it to the user-subchannel matching problem.


\subsubsection{Matching Problem Formulation for the Tx-Rx Selection and Time Scheduling}

We formulate the Tx-Rx selection and time scheduling problem as a two-sided \emph{matching} between the set of users and time slots. If a user is matched with one time slot, we say that it acts as a Tx user in this time slot; otherwise it acts as a Rx user in this time slot. Denote the set of users as $\cal{N}$ and the set of time slots in one transmission period as $\cal{T}$. Mathematically, a \emph{matching} $\Psi$ is defined as a mapping from the set $\cal{N} \cup \cal{T} \cup \varnothing$ into itself such that $\Psi \left( {{j}} \right) \subseteq \mathcal{T}$, $j \in \cal{N}$ and $\Psi \left( {i} \right) \subseteq \mathcal{N}$, $i \in {\cal{T}}$. The number of scheduled Tx users in time slot $i$ is denoted as $\left| {\psi \left( {{i}} \right)} \right|$ and the number of time slots in which user $j$ acts as a Tx user is denoted as $\left| {\psi \left( {{j}} \right)} \right|$ satisfying $\left| {\psi \left( {{j}} \right)} \right| \le T_{max}$.

Any two users matching to the same time slot are defined as \emph{matching peers}, which is an expansion of ``roommates" in the SR problem. A pair of users $\left( j, {j'} \right)$ is defined as a \emph{forbidden pair} in time slot $i$ if $d_{j,j'}^{(i)} \le r$, that is, they cannot be matched with this time slot simultaneously due to constraint $\left( 7a \right)$ as mentioned above\footnote{Note that we have modified the definition here to better fit our scenario, which differs from the traditional definition of forbidden pair in~\cite{7AM-1992}.}.
%

\emph{a) Preference relation:} To better describe the interaction between the users, we then investigate how each user selects its matching peers. The key idea is that each user tends to choose those far away from it as its matching peers so that the overlapping part of their communication disks will be small and the number of conflicting Rx users will decrease, thereby reducing the potential collision. For convenience, we call such potential collision as cross influence brought by multiple Tx users on the conflicting Rx users. Therefore, it is natural that the cross influence on the Rx users brought by any two Tx users $j$ and $j'$ in time slot $i$ can be evaluated by the overlapping area of their communication disks, as shown below:
\begin{equation} \label{cross_influence_area}
I_{j,j'}^{(i)} =
\begin{cases}
{\left( {2r - d_{j,j'}^{(i)}} \right)^2}, &~ \mbox{if $2r > d_{j,j'}^{(i)}$},\\
\varepsilon, &~ \mbox{otherwise,}
\end{cases}
\end{equation}
in which $-0.1 < \varepsilon < 0$ is small enough. Therefore, the average cross influence brought by user $j$ in time slot $i$ is then given by
\begin{equation} \label{user_time_utility}
Q_j^{(i)} =
\begin{cases}
\frac{1}{\Psi \left( {{i}} \right) + 1}{\sum\limits_{j' \in \Psi \left( {{i}} \right)} I_{j,j'}^{(i)}}, &~ \mbox{if $\left| {\psi \left( {{i}} \right)} \right| > 1$},\\
\varepsilon, &~ \mbox{if $\left| {\psi \left( {{i}} \right)} \right| = 1$}.
\end{cases}
\end{equation}
Given any two time slots $i$ and ${i'}$, user $j$ \emph{prefers} $i$ to ${i'}$ if user $j$ causes smaller cross influence in time slot $i$ and no forbidden pair exists when it matches with time slot $i$, i.e., $Q_j^{(i)} < Q_{j'}^{(i)}$ and $d_{j,j'}^{(i)} > r,\forall j' \in \Psi \left( {{i}} \right)$. For convenience, we denote the \emph{preference relation} as ${i}{ \succ _{{j}}}{{i'}}$. The total cross influence with respect to time slot $i$ is the sum of the cross influence of all users matched with time slot $i$, which is omitted here.

\emph{b) Analysis on matching formulation:} Based on equations $\left( \ref{cross_influence_area} \right)$ and $\left( \ref{user_time_utility} \right)$, the matching between users and time slots is a multi-dimensional (MD) geometric SR problem with forbidden pairs, which is different from the traditional 2D or 3D SR problem due to the extension of multiple dimension and forbidden pairs. In a traditional geometric SR problem, any user sets its preference based on the distance to other users, and a time slot can only be occupied by two or three users. A stable matching can be established via a greedy algorithm such that no two/three users prefer each other to their current matches. However, in such a stable matching, the users who are closest to each other always match with the same time slot~\cite{ESAKJV-2009}. This may not be suitable for the case with forbidden pairs since the set of closest matching peers always contains forbidden pairs with the smallest distance, making the stable matching an infeasible solution.

If there is no matched forbidden pair in the final matching, we then say that it is a \emph{feasible solution}. The following proposition, proved in Appendix B, will justify whether a feasible solution exists in our case.

\textbf{Proposition 1:} A feasible solution exists given a MD-geometric SR problem if
\begin{equation} \label{colorable}
\mathop {\max }\limits_{\mathcal{H},i} \mathop {\min }\limits_{j \in \mathcal{H}} \left| {{\mathcal{F}}_j^{(i)}} \right| + 1 \le {T_v},
\end{equation}
in which $\mathcal{H} \subseteq \mathcal{N}$ and ${\mathcal{F}}_j^{(i)} = \left\{ {j' \in \mathcal{H}|d_{j,j'}^{(i)} \le {r}} \right\}$ is the subset of users with the distance smaller than $r$ from user $j$.


\subsubsection{Rotation Matching Algorithm Design for the MD-geometric SR Problem}

A two-phase matching algorithm is designed in which a feasible solution is obtained in the first phase, and then the users rotate their matching peers to further reduce their average cross influence while avoiding the forbidden pairs in the second phase.

\emph{a) Phase 1 -- Obtaining a feasible solution:}  We utilize a greedy algorithm in which each user is assigned with one time slot if the matching problem is solvable. For any unmatched user $j$, if it can form a forbidden pair with user ${j'} \in \Psi(i)$, user $j$ is not allowed to match with time slot $i$. Each user $j$ searches for the set of available matched time slots obtained by
\begin{equation} \label{avaible_time}
{\mathcal{P}_j} = \left\{ {{i} \in {\mathcal{T}}_{matched}|\Psi \left( {{i}} \right) \cap {\mathcal{H}}_j^i = \varnothing } \right\},
\end{equation}
in which ${\mathcal{T}}_{matched}$ is the set of all matched time slots in current matching. When user $j$ is not allowed to match with all the matched time slots, i.e., ${\mathcal{P}_j} = \varnothing$, it then selects an unmatched time slot $i'$, i.e., $\Psi \left( {{{i'}}} \right) = \varnothing$. Otherwise, it selects a matched time slot ${i^*}$ such that it brings the smallest cross influence $Q_j^{(i)}$, i.e.,
\begin{equation} \label{utility_timeS}
{T_{(i^*)}} = \mathop {\arg }\limits_i \min Q_j^{(i)}.
\end{equation}
The whole process continues until all users or time slots are matched.

\emph{b) Phase 2 -- Rotation matching:} We introduce the concept of \emph{rotation} to better describe the interdependency of different users in the MD-geometric SR problem.

\textbf{Definition 1:} Given a matching $\Psi$ obtained from the first phase with a subset of users ${\mathcal{N}_{s}} \subseteq \cal{N}$ satisfying $\left| {{{\cal N}_s}} \right| = L \ge 2$, a \emph{rotation sequence} refers to
\begin{equation} \label{rotation_sequence}
\zeta_l  = \left( {{N_s(1)},\Psi \left( {{N_s(l+1)}} \right)} \right),\left( {{N_s(2)},\Psi \left( {{N_s(l+2)}} \right)} \right), \cdots ,\left( {{N_s(L)},\Psi \left( {{N_s(l)}} \right)} \right),
\end{equation}
where $1 \le l \le L-1$ implying that there exists $L-1$ possible rotation sequences given matching $\Psi$ and user subset ${\mathcal{N}_{s}}$. When $l = 1$, $\zeta_L$ refers to the original matching sequence in $\Psi$. A \emph{rotation matching} $\Psi _{{\mathcal{N}_s}, \zeta_l}^{rot}$ w.r.t. rotation sequence $\zeta_l$ is defined as
\begin{equation}
\Psi _{{\mathcal{N}_s}, \zeta_l}^{rot} = \Psi \backslash \left\{ {\left( {{N_s(1)},\Psi \left( {{N_s(1)}} \right)} \right), \cdots ,\left( {{N_s(q)},\Psi \left( {{N_s(q)}} \right)} \right)} \right\} \cup \zeta_l.
\end{equation}

To be specific, a rotation matching is generated in which a subset of users switch their matches in a pre-defined cyclic order shown in $\left( \ref{rotation_sequence} \right)$ while other users remain their matches. At least one user/time slot in the rotation sequence is allowed to be unmatched, i.e., it is matched with a dummy time slot or a dummy user.

Aiming at reducing the cross influence on the unmatched users (i.e., Rx users) brought by the matched users (i.e., Tx users), we evaluate the optimality of one rotation matching as well as its validity as shown below.

\textbf{Definition 2:} Consider a matching $\Psi$ and a subset of users ${\mathcal{N}_{s}} \subseteq \cal{N}$ with size $L$. For $1 \le l \le L-1$, a rotation matching $\Psi _{{\mathcal{N}_s}, \zeta_l}^{rot}$ is \emph{valid} if any user $j \in {\mathcal{N}_s}$ does not form forbidden pairs with any matching peers in $\Psi _{{\mathcal{N}_s}, \zeta_l}^{rot}\left[ {\Psi _{{\mathcal{N}_s}, \zeta_l}^{rot}\left( {{j}} \right)} \right]\backslash \left\{ {{j}} \right\}$. A rotation matching is \emph{optimal} if it achieves the smallest average cross influence in $\left( \ref{user_time_utility} \right)$ with no forbidden pairs among the matching peers, i.e., ${l ^{opt}} = \mathop {\arg }\limits_{1 \le l \le L}  \min {\sum _{j \in {{\cal N}_s}}}{\sum _{i \in \Psi _{{\mathcal{N}_s}, \zeta_l}^{rot}\left( {{j}} \right)}}Q_j^{(i)}$.

Note that in Definition 2, we have extended the optimality of a rotation matching to include the case of $l = L$ so as to compare the original matching sequence with the rotation sequences.

\textbf{Remark 2:} Two special cases of the rotation matching are discussed as below.

(1) When $q = 2$, the rotation matching is degraded to a swap matching as defined in~\cite{BLCHW-2011,BLY-2016}, and the dummy nodes are valid in the rotation matching.

(2) For users $j, {j'} \in {\cal{N}}_s$, if $q > 2$, ${\Psi \left( {{j}} \right)}$ and ${\Psi \left( {{j'}} \right)}$ are allowed to be the same; otherwise, there exists no rotation.

%

Note that the concept of rotation matching describes a more general case compared to the swap matching~\cite{BLCHW-2011}, where only two users exchange their matches for utility improvement. It provides a heuristic method to cope with the interaction of the players in the SR problem. To achieve a tolerable complexity of the algorithm, we set the length of each rotation sequence $q \le q_{max}$.


Based on the defined rotation matching, we then define the stability as shown below.

\textbf{Definition 3:} A MD-geometric SR matching $\Psi$ is \emph{$q$-exchange stable} if all rotation matchings in $\Psi$ are optimal, i.e., no rotation matchings with $q \leq q_{max}$ can further reduce the average cross influence of $\Psi$.


%
%
%
%
%
\emph{c) Description of the whole algorithm:}
Given the above definitions of the rotation and stability, we then propose a Tx-Rx selection and time scheduling algorithm (UTSA) to obtain a $q$-exchange stable matching, as listed in detail in Table \ref{alg-UT}.
\begin{table}[!t]
\renewcommand{\arraystretch}{1.0}
\caption{Tx-Rx Selection and Time Slot Allocation Algorithm (UTSA)}
\label{alg-UT}
\centering
\begin{tabular}{p{150mm}}
\hline
\textbf{Input:} Set of users $\mathcal{N}$; Set of time slots $\mathcal{T}$; locations of all the users.

\textbf{Output:} A $q$-exchange stable matching $\Psi$ in which the set of Tx users and Rx users are determined for each time slot.

\textbf{Initialization}

\quad Record current matching as $\Psi$, and construct the list $\mathcal{T}_{matched} = \{ 1, 2, \cdots, T_v\}$.

\quad For each time slot $i$, each user $j$ constructs the forbidden pair list $\mathcal{H}_j^{(i)}$.

{\bf Phase 1: \emph{Obtaining a feasible matching}}

\quad \textbf{for} $j = 1: N$,

\quad \quad Obtain the available set of matched time slots as ${\mathcal{P}}_j$ according to $\left( \ref{avaible_time} \right)$.

\quad \quad \textbf{if} ${\mathcal{P}_j} = \varnothing$ and $|{\mathcal{T}_{unmatched}}| \ne \varnothing$

\quad \quad \quad Randomly select a time slot $i^*$ from ${{\cal{T}}_{unmatched}}$;

\quad \quad \textbf{else}

\quad \quad \quad Obtain ${T_{i^*}}$ according to $\left( \ref{utility_timeS} \right)$.

\quad \quad \textbf{end}

\quad \textbf{end}


\quad Set $\Psi \left( {{j}} \right) = {i^*}$;

\quad Update ${\mathcal{T}_{unmatched}} = {\mathcal{T}_{unmatched}}\backslash \left\{ {{i^*}} \right\}$.


\textbf{Phase 2: \emph{Rotation Matching}}

\quad \textbf{Repeat:}

\quad \quad Randomly select a user sequence $\mathcal{N}_s$;

\quad \quad Find the optimal rotation matching ${\ell ^{opt}}$ of $\mathcal{N}_s$ according to Definition 2;

\quad \quad Update $\Psi  = \Psi _{{\mathcal{N}_s},{\ell ^{opt}}}^{rot}$;

\quad \textbf{until} no rotation matching can further reduce the average cross influence to the network.

\textbf{\emph{End of algorithm.}}\\
\hline

\end{tabular}
\end{table}
To start with, the BS collects each user's geometric information to construct the forbidden matching list of each user. In Phase 1, a feasible matching is obtained by the BS via a greedy algorithm in which each user is matched with exactly one time slot, i.e., each user acts as a Tx user in its matched time slot and acts as a Rx user in the other time slots. Phase 2 contains multiple iterations in each of which an optimal rotation matching is executed by the BS given a randomly selected rotation sequence. The iteration stops until no rotation sequence can further reduce the total cross influence of the network. The UTSA is performed by the BS for centralized resource allocation.

\subsubsection{Multi-dimensional Stable Roommate Matching for the Sub-channel Allocation}

We now extend the above algorithm to solve the sub-channel allocation problem in each time slot. We re-construct the utility function based on $\left( \ref{utility} \right)$ to formulate a non-convex integer programming problem, which is much more intractable than the time-user matching problem.

\emph{a) Definitions:} We observe that the set of Tx users and subchannels can be considered as ``students" and ``rooms" in a MD SR problem, respectively. Denote the set of Tx users in each time slot $i$ as $\mathcal{N}_t^{(i)}$, and the set of subchannels as $\cal{K}$. In each time slot $i$, a \emph{user-subchannel matching} $\Phi_{i}$ is then defined as a mapping from the set $\mathcal{N}_{t}^{(i)} \cup {\mathcal{K}} \cup \varnothing$ into itself such that $\Phi \left( {{j}} \right) \subseteq {K}$ and $\Phi \left( {k} \right) \subseteq {\mathcal{N}_{t}^{(i)}}$, in which $j \in {\mathcal{N}_{t}^{(i)}}$, $k \in \cal{K}$. Each Tx user can match with multiple sub-channels and each sub-channel can match with multiple Tx users. To be specific, we have $\left| {\Phi \left( k \right)} \right| \le {K_u}$ and $\left| {\Phi \left( {{j}} \right)} \right| \le {K_{\max }}$ based on constraints $\left( 7b \right)$ and $\left( 7d \right)$.


\emph{b) Preference and utility functions:} Different from the geometric utility function in $\left( \ref{user_time_utility} \right)$, we further evaluate the co-channel cross influence based on the number of successfully decoded signals, which is our final goal of maximization. The utility functions of the Tx users and sub-channels need to be re-defined in this case.

For any Tx user $j$, its utility with regard to subchannel $k$ represents the approximation of the number of successfully decoding Rx users in its communication disk, i.e., ${\sum _{m \in {\cal N}_j^{\left( i \right)}}}U_{j,m,k}^{\left( i \right)}$. If Tx user $j$ obtains higher utility from subchannel $k$ than from subchannel ${k'}$, we have $k{ \succ _j}k'$. Similarly, for any subchannel $k$, its preference ${ \succ _{{k}}}$ over the subset of Tx users $\mathcal{N}_t^{(i)}$ and $\mathcal{N}_{t'}^{(i)}$ is defined based on the utility function as below. Given two matchings $\Phi$, $\Phi'$, and $\Phi (k) = \mathcal{N}_t^{(i)}$, $\Phi' (k') = \mathcal{N}_{t'}^{(i)}$, we have
\begin{equation} \label{preference_subchannel}
\left\{ {\mathcal{N}_t^{(i)},\Phi } \right\}{ \succ _{k}}\left\{ {\mathcal{N}_{t'}^{(i)},\Phi '} \right\} \Leftrightarrow \sum\nolimits_{{j} \in \mathcal{N}_t^{(i)}} {\sum\nolimits_{m \in N_j^{(i)}} {U_{j,m,k}^{(i)}} }  > \sum\nolimits_{{j} \in \mathcal{N}_{t'}^{(i)}} {\sum\nolimits_{m \in N_j^{(i)}} {U_{j,m,k'}^{(i)}} }.
\end{equation}
Due to the symmetry of the preference relation of the users and sub-channels, the benefit of subchannel $k$ matching with Tx user $j$ is higher than that of subchannel ${k'}$ matching with Tx user $j$ if Tx user $j$ prefers subchannel $k$ to subchannel ${k'}$, i.e., $k{ \succ _j}k'$.

Given a subset of users $\cal{N}_s$, the optimal rotation matching in time slot $i$ can be rewritten as,
\begin{equation} \label{optimal_rotation}
{\ell ^{(i, opt)}} = \mathop {\arg }\limits_\ell  \max \sum\nolimits_{j \in {\mathcal{N}_t^{(i)}}} {\sum\nolimits_{m \in \mathcal{N}_j^{(i)}} {\sum\nolimits_{k \in \Phi _{{\mathcal{N}_s},\ell }^{rot}\left( {{j}} \right)} {U_{j,m,k}^{(i)}} } }
\end{equation}
With the above definitions, the rotation matching algorithm proposed in Section \uppercase\expandafter{\romannumeral3}.B.2 is then modified as shown in Table \ref{alg-US}. For each time slot, the matching algorithm is performed and stops when the system utility cannot be further improved by any rotation matchings.

The whole centralized Tx-Rx selection and time-frequency resource allocation scheme is then obtained after the BS performs the above two matching algorithms in Section \uppercase\expandafter{\romannumeral3}.B.2 and \uppercase\expandafter{\romannumeral3}.B.3.

\begin{table}[!t]
\renewcommand{\arraystretch}{1.0}
\caption{Rotation Matching Algorithm for Sub-channel Allocation (RMSA)}
\label{alg-US}
\centering
\begin{tabular}{p{150mm}}
\hline

\textbf{Initialization:}

\quad Set ${\mathcal{L}_{record}} = \varnothing$;

\quad Each Tx user $j$ randomly matches with a subset of subchannels with the size smaller than ${K_{\max }}$.

\textbf{Repeat:}

\quad Select a user subset ${\mathcal{N}}_s$ and the corresponding rotation sequence $\ell \notin \mathcal{L}_{record}$;

\quad \textbf{if} $\ell$ is not optimal

\quad \quad Set $\ell^{opt} = \varnothing$;

\quad \quad \textbf{while} $\ell^{opt} = \varnothing$

\quad \quad \quad Obtain ${\ell ^{temp}}$ based on equation $\left( \ref{optimal_rotation} \right)$;

\quad \quad \quad \textbf{if} the updated matching $\Phi _{{{\mathcal{N}}_s, \ell ^{temp}}}^{rot}$ satisfies constraints $\left( \ref{system_3} \right)$ and $\left( \ref{system_5} \right)$

\quad \quad \quad \quad Set ${\ell ^{opt}} = {\ell ^{temp}}$;

\quad \quad \quad \quad Add ${\ell ^{opt}}$ to $\mathcal{L}_{record}$;

\quad \quad \quad \quad \textbf{break};

\quad \quad \quad \textbf{else}

\quad \quad \quad \quad Remove ${\ell ^{temp}}$ from the candidate set $\{ \ell \}$.

\quad \quad \quad \textbf{end}

\quad \quad \textbf{end}

\quad \textbf{end}

\textbf{until} no rotation matching can further improve the total utility of the network in equation $\left( \ref{system_1} \right)$.

\textbf{\emph{End of algorithm.}}\\
\hline

\end{tabular}
\end{table}

\section{NOMA-based Distributed Power Control and Algorithm Design}

To perform the SIC decoding, necessary prior knowledge needs to be provided to the Rx users, e.g., the CSI of the intended Tx user -- Rx user links, the number and transmit power of corresponding Tx users. Therefore, the distributed power control requires information exchange between the Tx users and Rx users, i.e., the \emph{control signaling}, as will be explained below.

%
%
\subsection{Control signaling}
We divide each time slot\footnote{We assume that all the users are synchronized to a common timing source, i.e., the BS, supported by the LTE system.} into one control signaling portion and one data transmitting portion, in which the control portion is divided into several blocks\footnote{Each block is set for reference signal transmission with the duration of less than 200 $\mu$s.} for control message exchange between the Tx users and Rx users. 

\begin{figure}[!t]
\centering
\includegraphics[width=4in]{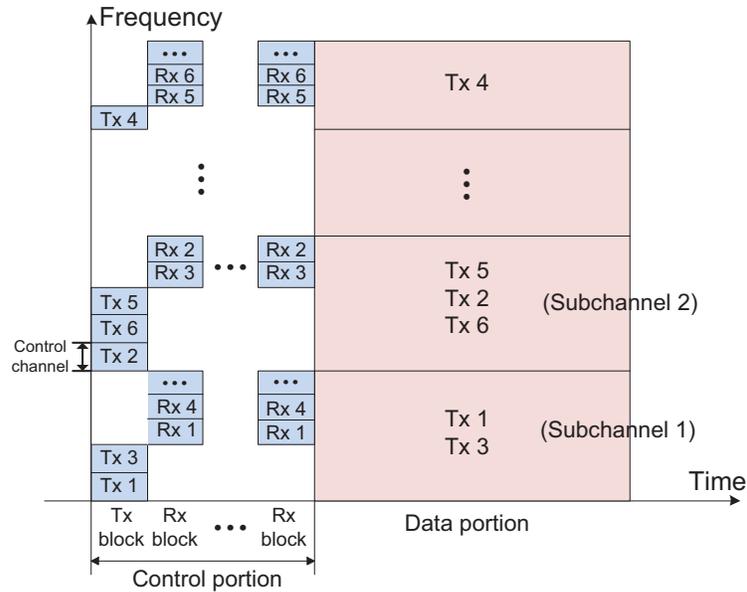}
\caption{The user association process within one time slot.} \label{control}
\end{figure}

The structure of one time slot is shown in Fig.~\ref{control}. Based on the subchannel allocation of the data portion, each subchannel may be allocated to multiple Tx users, e.g., Tx user 1 and 3 occupies subchannel 1 in Fig.~\ref{control}. For the control portion consisting of $T_c$ pairs of Tx blocks and Rx blocks, each Tx user $j$ and Rx user $m$ is assigned a unique frequency position in the Tx block and Rx block, respectively. For example, two positions in subchannel 1 are assigned to Tx user 1 and Tx user 3 with smaller bandwidth of channels such that the CSI of the data channels can be correctly measured by the control signals within the coherent bandwidth. The remained parts of subchannel 1 are allocated to the Rx users in the neighborhood of Tx user 1 and Tx user 3, e.g., Rx user 1, Rx user 4, and so on.

Each pair of a Tx block and a Rx block work as below.
\begin{itemize}
\item \textbf{Tx block:} every Tx user $j$ broadcasts a reference signal to its neighborhood containing the transmit power and the frequency resources that it will occupy for the data transmission. The transmitted power $p_j$ is determined based on the feedback sent by the Rx users in previous block. Each Rx user $m$ obtains its neighboring Tx users' CSI via the received reference signals.
\item \textbf{Rx block:} each Rx user $m$ evaluates whether it will successfully decode the coming data messages based on the information obtained from the Tx block. It then calculates the potential co-channel interference to each Tx user $j$ caused by other Tx users, and broadcasts the feedback for further processing in the next Tx block.
\end{itemize}


\subsection{Power control problem for each Tx user}
We now describe how Tx user $j$ decides its transmitted power in each Tx block of time slot $i$. For convenience, we denote the set of Rx users sending feedback to Tx user $j$ during the Rx block over subchannel $k$ in time slot $i$ as $\mathcal{B}_{j,k}^{(i)}$.

Note that in a dense V2X broadcast system, it is very hard that each Tx user $j$'s signals can be successfully decoded by all neighboring Rx users. Therefore, we set a unique weighted value $w_j \in (0, 1]$ for each Tx user $j$ such that at least $100w_j$ percent of the Rx users in the neighborhood should be able to decode the signal of Tx user $j$. The weighted value is updated once by the BS at the beginning of each SPS period. The power control problem of Tx user $j$ over subchannel $k$ in each Tx block of time slot $i$ can then be formulated as below:
\begin{subequations}\label{power_control}
\begin{align}
& \mathop {\min } \quad p_{j,k}^{(i)} \label{tx_1}\\
s.t.~ & \sum\limits_{m \in \mathcal{B}_{j,k}^{(i)}} \left[ {R_{j,m,k}^{(i)}({p_{j,k}^{(i)}}) - {{\bar R}_{th}}} \right]^+  \ge {w_j}\left| {\mathcal{B}_{j,k}^{(i)}} \right|, \label{tx_2}\\
& p_{j,k}^{(i)} \le P \label{tx_3},
\end{align}
\end{subequations}
in which $P$ is the maximum transmitted power of Tx user $j$ over subchannel $k$. In problem $\left( \ref{power_control} \right)$, Tx user $j$ aims to transmit with the minimum power such that at least $100w_j$ percent Rx users in $\mathcal{B}_{j,k}^{(i)}$ can decode their signals. In this way, the cross influence caused by multiple Tx users can be constrained to a tolerable level. Problem $\left( \ref{power_control} \right)$ can be solved by utilizing a bisection method~\cite{JR-1983}. The whole distributed power control and user association process is presented in detail in Table~\ref{alg-DP}.

\begin{table}[!t]
\renewcommand{\arraystretch}{1.0}
\caption{Control Portion in the Distributed Power Control of the Users in time slot $i$}
\label{alg-DP}
\centering
\begin{tabular}{p{150mm}}
\hline

\textbf{\emph{Initialization:}} each user is assigned a unique frequency position for transmission in the control portion.

\textbf{for} $t = 1: T_c$

\quad \textbf{Tx block:}

\quad \quad \textbf{if} $t = 1$

\quad \quad \quad Each Tx user broadcasts a reference signal with the initial transmit power $p_0^{(i)}$ over its allocated subchannels;

\quad \quad \textbf{else}

\quad \quad \quad Each Tx user $j$ obtains the optimal transmit power $p_{j,k}^{(i)}$ over subchannel $k$ by solving problem $\left( \ref{power_control} \right)$;

\quad \quad \quad It broadcasts a reference signal over subchannel $k$ with $p_{j,k}^{(i)}$.

\quad \quad \textbf{end}

\quad \quad Each Rx user $m$ receives the control signals of neighboring Tx users;

\quad \quad After decoding, Rx user $m$ knows the subset of subchannels assigned to each Tx user in the data portion;

\quad \quad Rx user $m$ obtains the CSI and transmitted power.

\quad \textbf{Rx block:}

\quad \quad \textbf{for} each Rx user $m$:

\quad \quad \quad Set ${F}$ as an empty matrix, and $l = 0$;

\quad \quad \quad \textbf{for} the received signal from Tx user $j$:

\quad \quad \quad \quad Set the subchannel occupied by Tx user $j$ in the data protion as $k$;

\quad \quad \quad \quad Calculate the potential interference item for Tx user $j$ over subchannel $k$ as shown in $\left( \ref{rate_equation} \right)$;

\quad \quad \quad \quad \textbf{if} Rx user $m$ can decode the coming data message of Tx user $j$

\quad \quad \quad \quad \quad Add [Tx user $j$, subchannel $k$, interference item] to ${F}(l)$;

\quad \quad \quad \quad \quad $l = l + 1$.

\quad \quad \quad \quad \textbf{else if} Rx user $m$ can decode the coming data message of all Tx user $j'$ with higher decoding priority than Tx suer $j$

\quad \quad \quad \quad \quad Add [Tx user $j$, subchannel $k$, interference item] to ${F}(l)$;

\quad \quad \quad \quad \quad $l = l + 1$.

\quad \quad \quad \quad \textbf{end}

\quad \quad \quad \textbf{end}

\quad \quad \quad Rx user $m$ broadcasts ${F}$ to the neighboring Tx users.

\quad \quad \textbf{end}


\textbf{\emph{End of the control portion.}}\\
\hline

\end{tabular}
\end{table}

\section{Analysis of the Proposed Matching Algorithms}
In this section, we analyze the proposed matching algorithms in terms of stability, convergence, complexity and optimality. Since the core of both the UTSA and RMSA is the rotation matching algorithm, we take the UTSA for example.


\subsection{Stability and convergence}
As proved in Appendix C, the proposed UTSA for the centralized Tx-Rx selection and time scheduling is guaranteed to converge to a final matching $\Psi_{\lambda^*}$. We then illustrate that it converges to a $q$-exchange stable matching.

\textbf{Theorem 1:} The final matching obtained by the UTSA is a $q$-exchange stable matching.
\begin{proof}
We know that the proposed UTSA converges to a final matching $\Psi_{\lambda^*}$. Based on the definition of the rotation matching, given any subset of users $\mathcal{N}_s$, the corresponding rotation sequence $\ell$ in $\Psi_{\lambda^*}$ is either the optimal one or the only valid one. Thus, the matches of $\mathcal{N}_s$ with respect to $\ell$ must be the best choice for users in $\mathcal{N}_s$, with the strategies of other users in $\mathcal{N}\backslash {\mathcal{N}_s}$ unchanged. Hence, the final matching is a $q$-exchange stable matching.
\end{proof}

The convergence of the RSMA can be guaranteed since the total utility increases after each rotation matching and there exists an upper bound of the total utility.


\subsection{Complexity}
Given the convergence of the UTSA, we now analyze the complexity of its two phases in terms of the number of possible rotation matchings. As proved in Appendix D, the number of rotation matchings to be considered in each iteration of Phase 2 in the UTSA is bounded by ${{N^{{q_{\max }}}}}$, with $q_{max} \ll N$. Given the total number of iterations $I_{iter}$, the complexity can then be expressed by $O\left( {I_{iter}{N^{{q_{\max }}}}} \right)$.

Note that $I_{iter}$ cannot be given in a closed form since it is not trivial to know at which iteration the total cross influence stops decreasing, which is common in the design of most heuristic algorithms~\cite{DWSH-2017, PR-1982}. To evaluate the convergence, we will show the distribution of the total number of executed rotation matchings in both algorithms in Fig.~\ref{alphatwoA}.

We consider a special case in which $q_{max} = N$, and the influence of forbidden pairs is evaluated as shown below.

\textbf{Corollary 1:} Given an initial matching $\Psi$ and $F$ forbidden pairs, at least $F\left[ {{2^{N - 1}} - 2 + \frac{1}{{\left\lfloor {N/2} \right\rfloor }}} \right]$ rotation matchings are not valid for the UTSA with $q_{max} = N$.
\begin{proof}
See Appendix E.
\end{proof}

\textbf{Corollary 2:} Given $q_{max} = N$ and $F$ forbidden pairs, the number of rotation matchings to be considered in each iteration is bounded by $\left( {N - F} \right) \cdot {2^{N - 1}} + 2F - \frac{F}{{\left\lfloor {N/2} \right\rfloor }}$ in each iteration of Phase 2 in the UTSA.

\quad \emph{Proof:} When $q_{max} = N$, the number of all possible rotation sequences can be calculated by $\sum\limits_{q = 1}^N {\mathop {qC}\nolimits_N^q }  = N \cdot {2^{N - 1}}$. According to Corollary 1, the number of rotation matchings can then be evaluated as
\begin{equation}
N \cdot {2^{N - 1}} - F\left[ {{2^{N - 1}} - 2 + \frac{1}{{\left\lfloor {N/2} \right\rfloor }}} \right] = \left( {N - F} \right)\cdot{2^{N - 1}} + 2F - \frac{F}{{\left\lfloor {N/2} \right\rfloor }}
\end{equation}

\subsection{Optimality}
After performing the algorithms proposed in Table \ref{alg-UT} and \ref{alg-US}, we can obtain a sub-optimal solution of the centralized Tx-Rx selection and time-frequency resource allocation problem, which is a natural result since the MD SR problem is NP-hard.

\textbf{Proposition 2:} All local minimum points of $\Psi$ in the user-time matching problem and the local maximum point of $\Phi$ in the user-subchannel matching problem correspond to the $q$-exchange stable matchings.
\begin{proof}
We take the UTSA for example. Suppose that the total cross influence of the matching $\Psi$ is a local minimum point. If it is not a $q$-exchange stable matching, then there must exist at least one rotation matching that can strictly reduce the cross influence of the network according to Definition 3, which is contradicted to the presupposition that $\Psi$ is a local minimum. Therefore, the matching $\Psi$ must be $q$-exchange stable.
\end{proof}

\section{Simulation Results}
\begin{figure}[!t]
\centering
\includegraphics[width=4.4in]{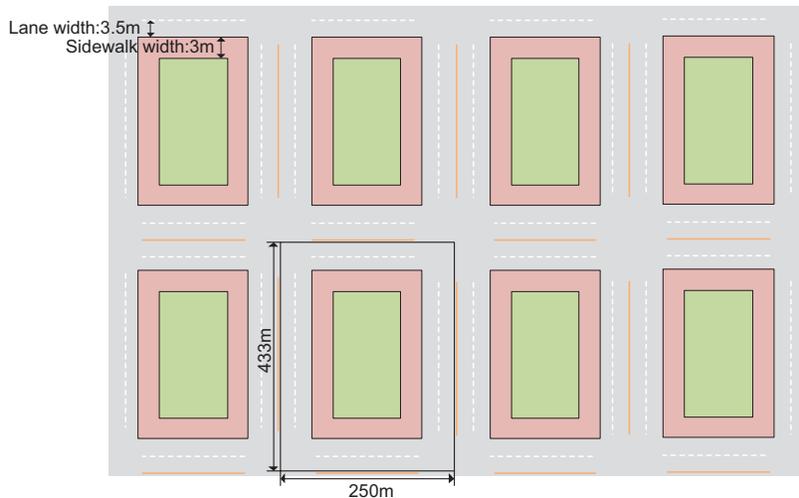}
\caption{Simulation setup for the road configuration in the urban scenario with the road width of 20$m$.} \label{partition}
\end{figure}

In this section, we consider the urban scenario defined in~\cite{3GPP-2016}, as shown in Fig.~\ref{partition}. The average inter-vehicle distance in the same lane is 2.5s$ \times v$, and the same density/speed in all the lanes is adopted in the simulation~\cite{3GPP-2016}. We evaluate the performance of the proposed NOMA-HCD scheme compared with the traditional OMA-based scheme and a NOMA-based geometric greedy algorithm (NOMA-GGA):
\begin{itemize}
\item \textbf{OMA-based scheme:} each Tx user transmits with the maximum power. The Tx-Rx selection and time scheduling is solved by utilizing a greedy algorithm, and the graph-based method in~\cite{CGMH-2014} is performed in the sub-channel allocation, in which the number of conflicting Rx users is minimized.
\item \textbf{NOMA-GGA:} we adopt the geometric utility function $\left( \ref{user_time_utility} \right)$ in the centralized SPS for both time and frequency resource allocation. A greedy algorithm proposed in phase 1 of Table \ref{alg-UT} is adopted.
\end{itemize}
For the simulations, we set each vehicle's peak power, $P_s$ to 23dBm, noise power spectral density to -174 dBm/Hz, carrier center frequency to 2GHz, system bandwidth to 10MHz~\cite{3GPP-2016}, decoding rate threshold to 2 bits/Hz/s, the number of considered time slots\footnote{Here we consider the scheduling period as 40ms for simplicity, and it can be easily extended to 100ms.} $T_v$ to 40, the length of each slot to 1 ms, the rotation matching parameter $q_{max}$ to 4, vehicle velocity to 15$\sim$60km/h, the size of each packet to 300 bytes, the ratio of control portion in one transmission slot to 15$\%$. The bandwidth is divided into 5 sub-channels and 10 sub-channels in the NOMA case and OFDMA case, respectively. The pass loss is obtained by a UMi model~\cite{3GPP-UMI}. Each active Tx user occupies one time slot in the worst case in each transmission period, implying that it needs to successively transmitting 300 bytes in this slot. Given the above parameters, the required transmission rates for meeting the low latency requirement is 1.41 bits/Hz/s in the NOMA case and 2.4 bits/Hz/s in the OMA case, respectively. Simulation results are obtained as shown below.

\begin{figure}
\centering
\subfigure[]{
\label{alphatwo:aA} 
\includegraphics[width=3.15in]{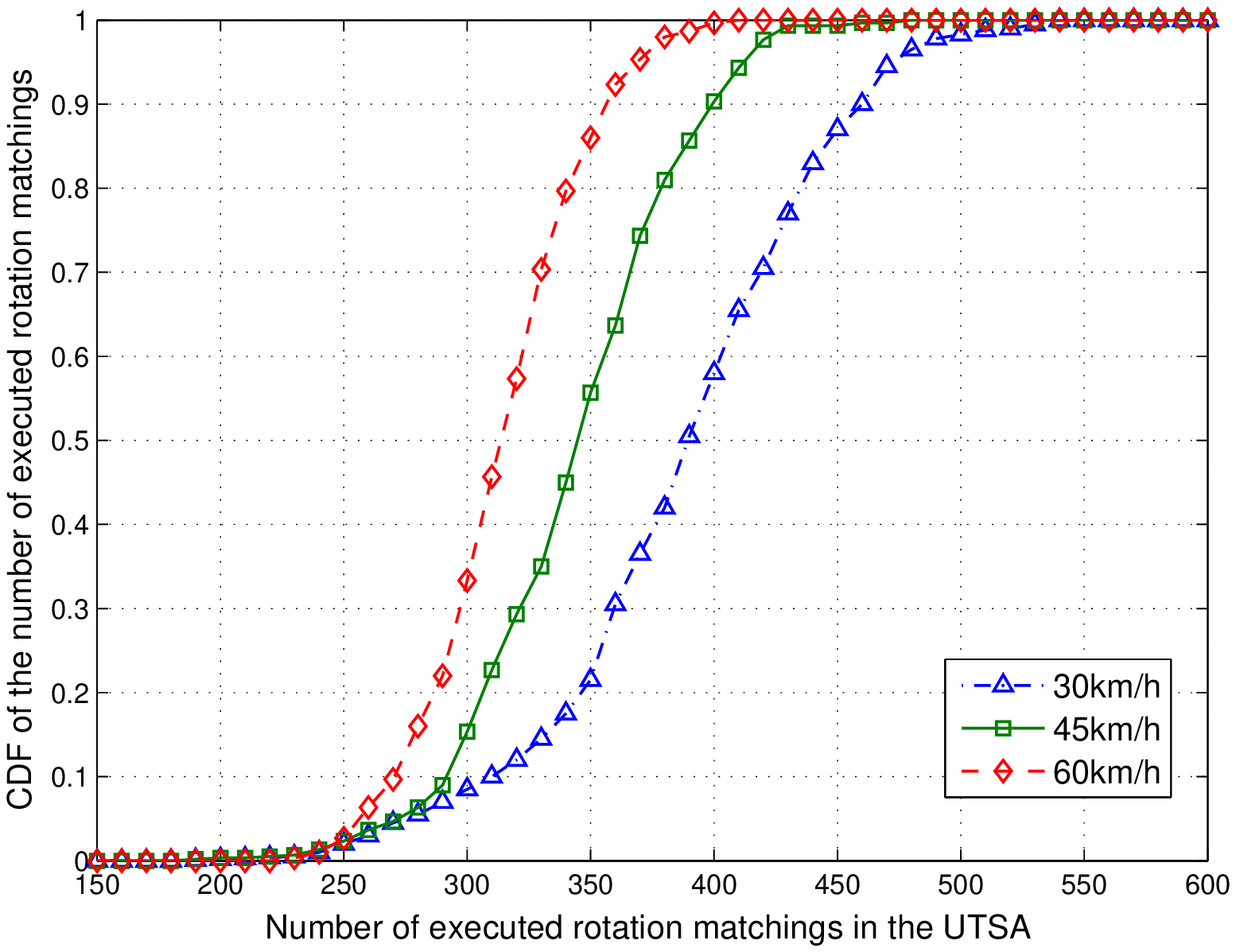}}
\hspace{-0.2in}
\subfigure[]{
\label{alphatwo:bB} 
\includegraphics[width=3.2in]{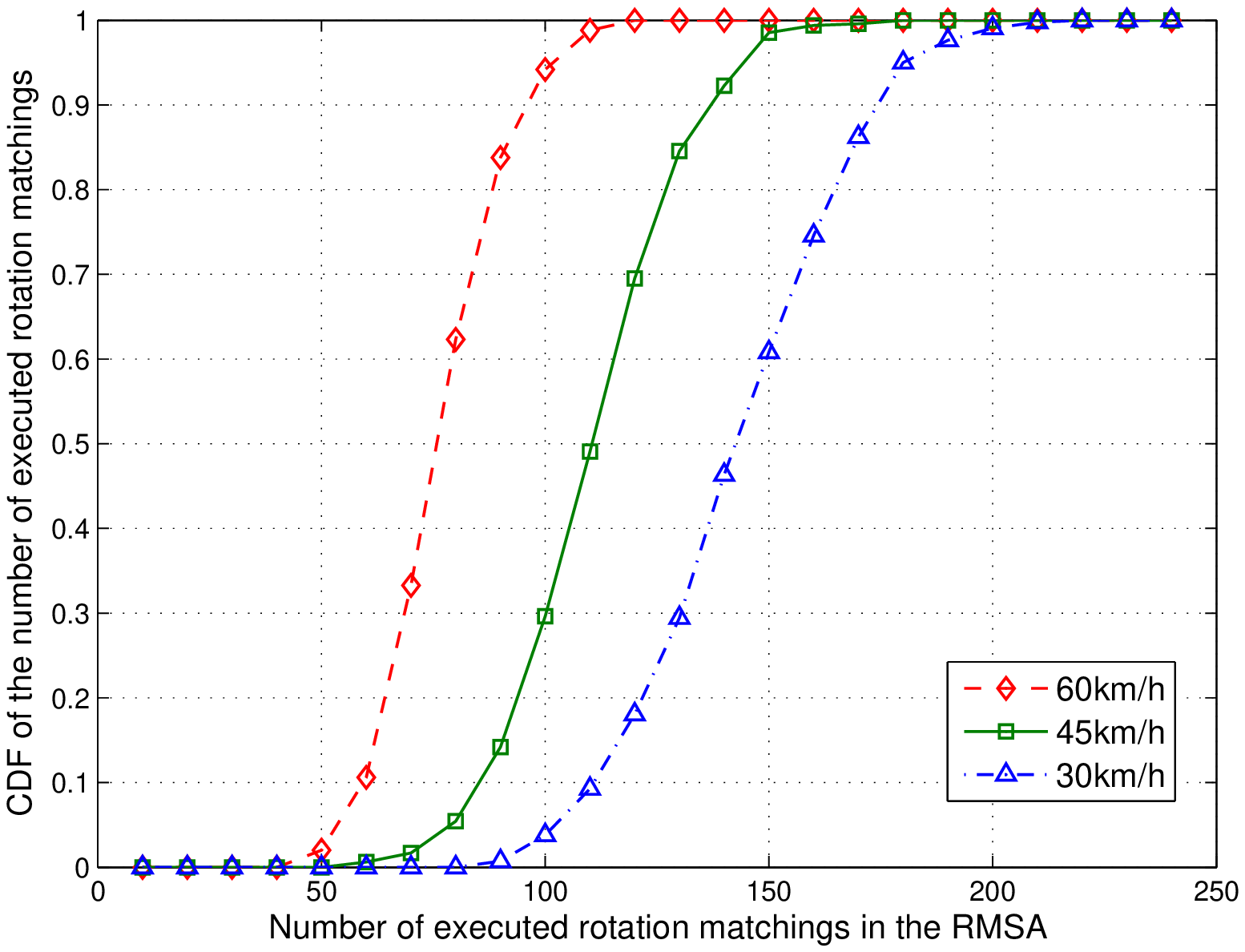}}
\caption{(a) C.D.F. of the total number of executed rotation matchings in the UTSA; (b) C.D.F. of the total number of executed rotation matchings in the RMSA.} \label{alphatwoA} 
\end{figure}

Let the random variable $\tilde Y$ denote the total number of executed rotation matchings $\lambda^*$ required for the proposed matching algorithms to converge. Fig.~\ref{alphatwoA} shows the cumulative distribution function (C.D.F.) of $\tilde Y$, $\Pr \left( {\tilde Y \le \tilde y} \right)$, versus $\tilde y$ for different velocity of vehicles in both algorithms, with the communication range of interest $r = 150$, and maximum number of users sharing the same sub-channel $K_{max} = 2$. We observe that due to the existence of the forbidden pairs, $\tilde Y$ in the UTSA is much smaller than the total number of potential rotation matchings as expressed in Proposition 7. Moreover, $\tilde Y$ in the RMSA is lower than that in the UTSA due to a smaller scale of the subchannel allocation problem.

\begin{figure}
\centering
\subfigure[]{
\label{alphatwo:a} 
\includegraphics[width=3.15in]{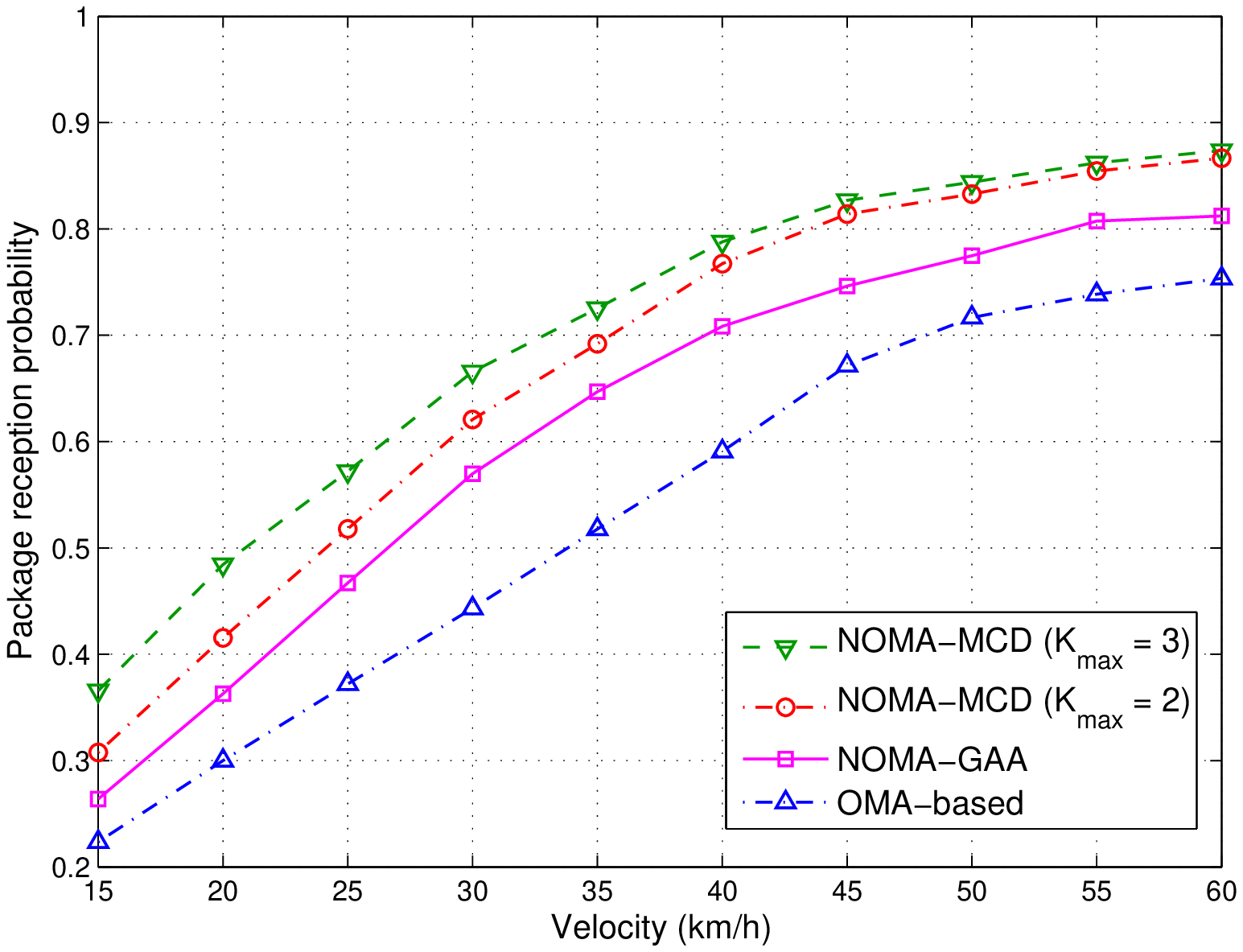}}
\hspace{-0.2in}
\subfigure[]{
\label{alphatwo:b} 
\includegraphics[width=3.2in]{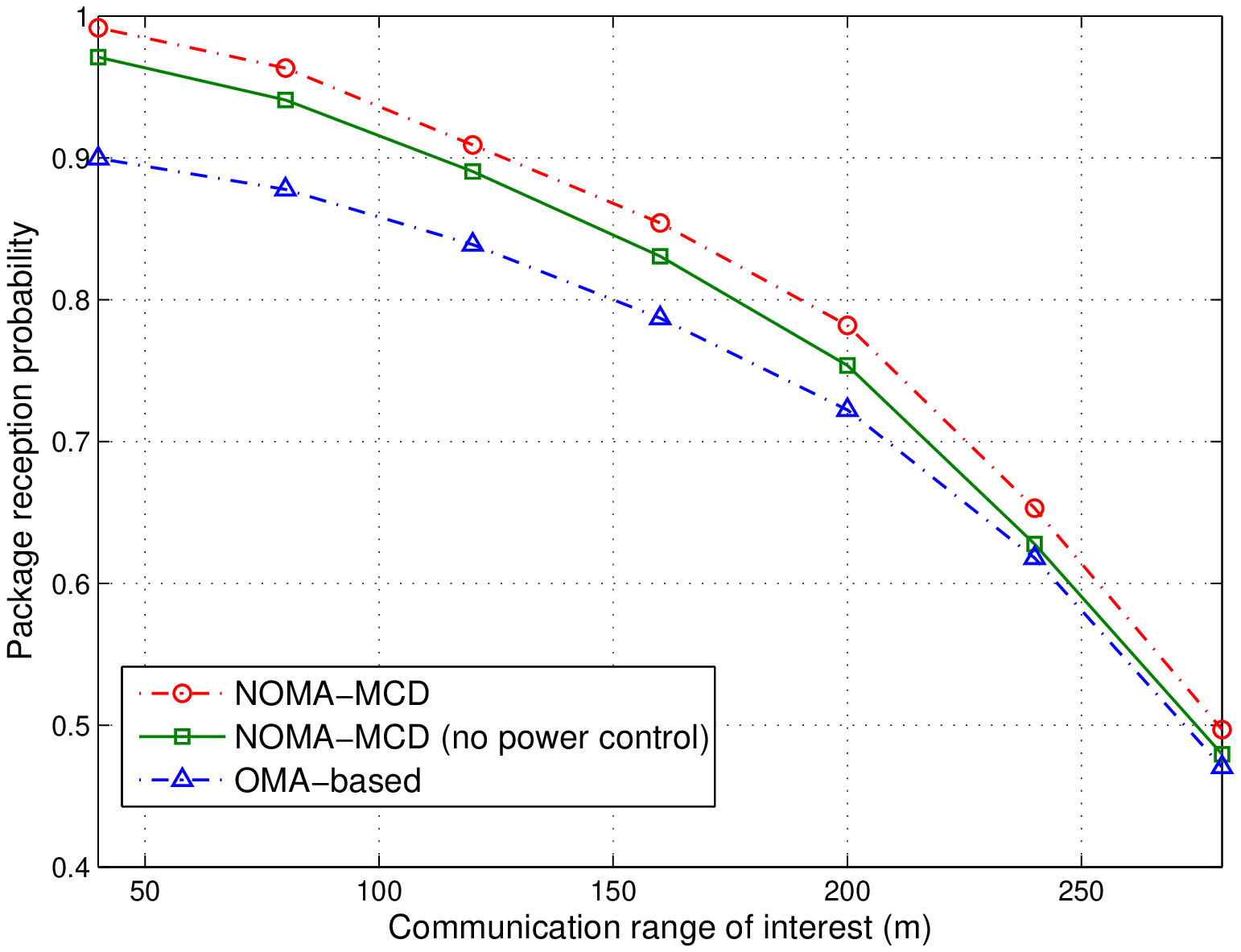}}
\caption{(a) PRP v.s. velocity of the vehicles; (b) PRP v.s. communication range of interest.} \label{alphatwo} 
\end{figure}

Fig.~\ref{alphatwo:a} illustrates the PRP v.s. the velocity of the vehicles, with the communication range of interest as 150m, and $K_{max}$ denoting the maximum number of Tx users sharing the same sub-channel. In the simulation, we consider a packet successfully received only when the latency requirement is also satisfied. We observe that the PRP increases with the velocity of vehicles, and the PRP growth becomes slower as $v$ grows. When the velocity grows, the density of users decreases in the network, and thus, there is less potential collision, leading to an increased PRP. Our proposed NOMA-MCD scheme performs better than the NOMA-GAA scheme and the OMA-based scheme due to a smart utilization of the time and frequency resources. Furthermore, the gap between the OMA-based scheme and the NOMA-MCD scheme becomes smaller as the velocity grows, indicating that the non-orthogonal manner of resource utilization works better in a dense network. Similarly, as $K_{max}$ becomes larger, more spectrum resources is available for the Tx users, especially when the density of vehicles is large, bringing a higher PRP.

%

Fig.~\ref{alphatwo:b} shows the PRP v.s. the communication range of interest $r$ with $K_{max} = 2$. The communication range of interest $r$ is set such that each Tx user only cares whether the Rx users within this range can receive its messages. It influences the formation of forbidden pairs in the UTMA, i.e., any two vehicles cannot act as Tx users simultaneously within the distance of $r$. As the range $r$ becomes larger, the number of active Tx users in one time slot decreases, i.e., the number of the accessed Tx users is smaller, leading to the declining of the average PRP of the network. Moreover, the gap between the NOMA-MCD scheme and the OMA-based scheme becomes smaller as $r$ increases, since the possibility of orthogonal use of the frequency resources is higher. Fig.~\ref{alphatwo:b} also shows that our proposed NOMA-MCD scheme performs better than the case without power control.


\begin{figure}
\centering
\subfigure[]{
\label{latency} 
\includegraphics[width=3.15in]{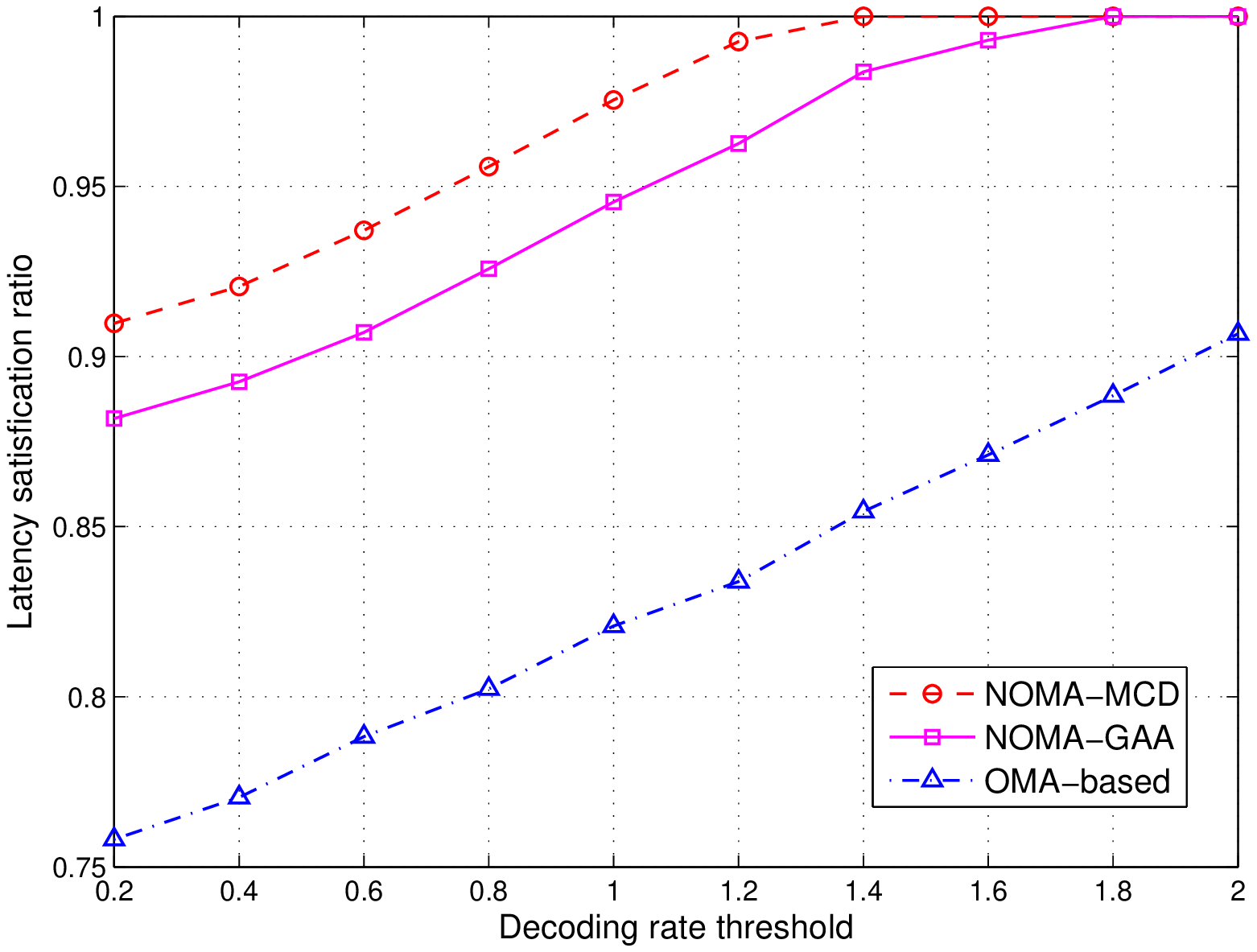}}
\hspace{-0.2in}
\subfigure[]{
\label{PRP_Kmax} 
\includegraphics[width=3.2in]{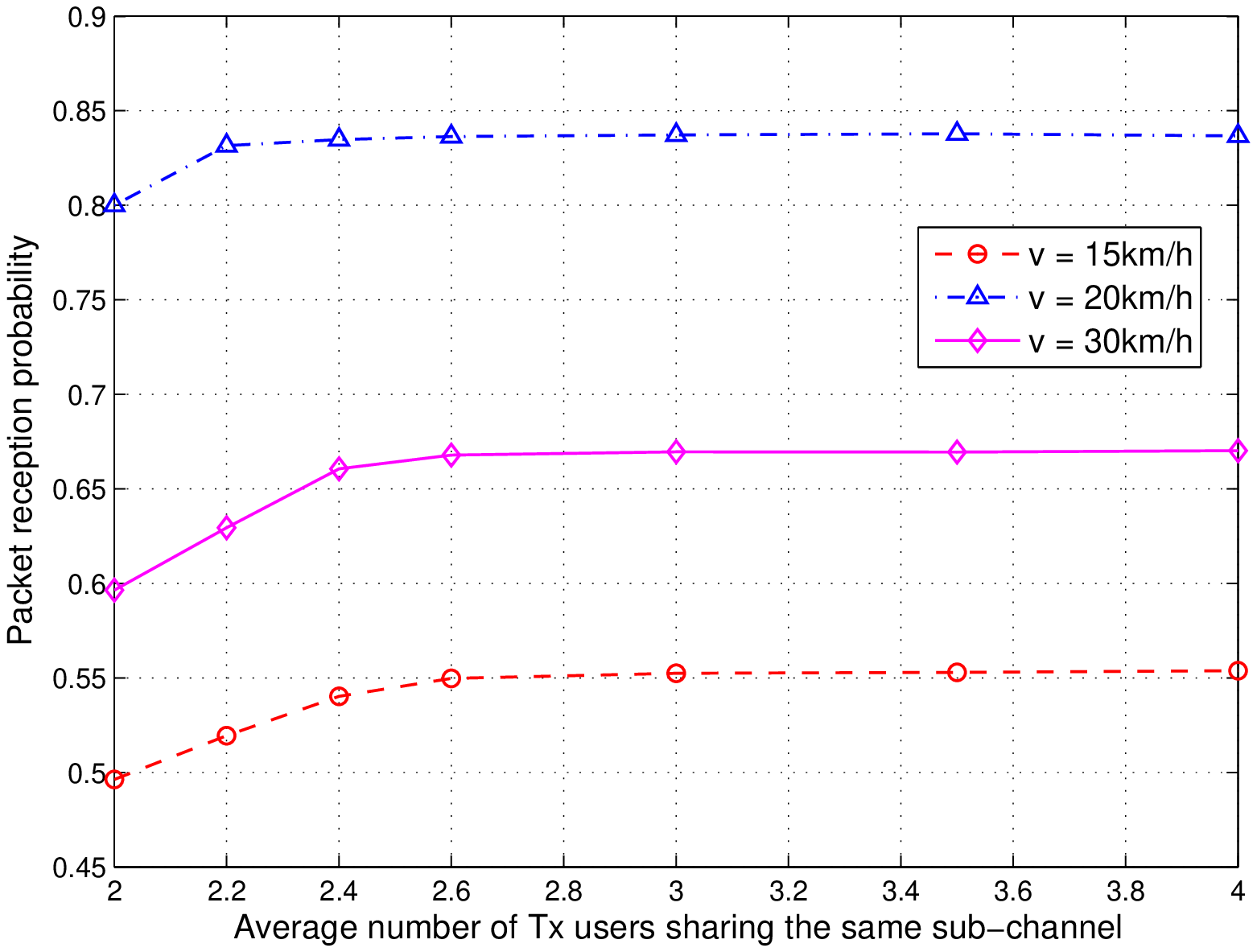}}
\caption{(a) Latency satisfaction ratio v.s. decoding rate threshold; (b) PRP v.s. the average number of users sharing the same sub-channel.} \label{doublelast} 
\end{figure}

Fig.~\ref{latency} presents the latency satisfaction ratio v.s. the decoding rate threshold ${{\bar R}_{th}}$ with $K_{max} = 2$. The latency satisfaction ratio is defined as the ratio between the number of successfully transmitted packets satisfying the latency requirement and the number of successfully decoded signals given the decoding rate (or SNR) threshold. The latency satisfaction ratio increases with the decoding rate threshold, since the decoding rate threshold becomes closer to required transmission rate for low latency. From Fig.~\ref{latency}, we observe that more Tx users can satisfy the latency requirement in the NOMA case compared to the OMA case. This is because the spectrum efficiency of the OMA case is lower than that of the NOMA case, leading to higher transmission delay.

Fig.~\ref{PRP_Kmax} depicts the PRP v.s. the average number of users sharing the same sub-channel, i.e., $K_{max}$, with the communication range of interest $r = 150$m. With different settings of the velocity $v$, the PRP grows to a stable value as $K_{max}$ increases, because each Tx user has more opportunity to access the channels. Considering the computational complexity of SIC which grows with $K_{max}$, we make an initial observation that when fewer than 3 Tx users share the same sub-channel, the balance between the spectrum efficiency and decoding complexity can be reached.


\section{Conclusions}
In this paper, we studied the scheduling and resource allocation problem in a cellular V2X broadcasting system. To reduce the access collision and improve the reliability of the network, we proposed a NOMA-based mixed centralized/distributed scheme, in which the time-frequency resources can be fully employed in a non-orthogonal manner by the BS, while the dynamic power control along with the iterative user association is performed by the vehicles to achieve a better performance of power domain multiplexing. To solve the centralized Tx-Rx selection and resource allocation problem, we re-formulated it as two MD-SR problems and proposed a novel rotation matching algorithm. Simulation results showed that the NOMA-MCD scheme outperforms the OMA-based scheme especially in a dense network.

\begin{appendices}
\section{Proof of NP-hardness of problem $\left( \ref{system_optimization} \right)$}
We prove that even the reduced version of $\left( \ref{system_optimization} \right)$ is a NP-hard problem, implying the original problem is also NP-hard~\cite{LP-2005}.

Given the Tx-Rx selection and time slot allocation, the original problem is then reduced to a sub-channel allocation problem in which the BS allocates sub-channels to the Tx users in each time slot such that the number of successfully decoded packets is maximized. We construct an instance of the reduced problem in which $K_u = 2$, and establish the equivalence between this instance and a vertex coloring problem~\cite{TB-1995}. Since the vertex coloring problem has been proved to be NP-hard, the instance of the reduced version of $\left( \ref{system_optimization} \right)$ is also NP-hard.

Consider the set of Tx users as nodes in a undirected conflicting graph. An edge exists between two nodes $j$ and $j'$ denoting they cannot be allocated the same sub-channel if either of the following conditions is satisfied: 1) Tx $j$ and $j'$ lie in each other's communication disk; 2) ${R_{j,m,k}} < {{\bar R}_j}$ or ${R_{j',m,k}} < {{\bar R}_{j'}}$, $\forall m \in \mathcal{N}_j^n \cap \mathcal{N}_{j'}^n$. Considering the set of sub-channels as colors, we aim at coloring as many nodes as possible such that no two adjacent vertices share the same color. For those dyed nodes, we say the corresponding Tx users can successfully send messages to the Rx users in their communication disks. Therefore, the equivalence between the vertex coloring problem and the instance is established.

A special case of problem $\left( \ref{system_optimization} \right)$ is NP-hard, and thus, the original problem in $\left( \ref{system_optimization} \right)$ is also NP-hard.

\section{Proof of Proposition 1}
Set $q = \mathop {\max }\limits_{{\cal H},i} \mathop {\min }\limits_{j \in {\cal H}} \left| {\mathcal{H}_j^{(i)}} \right|$. We need to prove that there exists a matching in which all users are matched, and any two matched peers do not form a forbidden pair given $(q+1)$ time slots. This can be achieved by constructing a sequence of users for matching. By setting $\mathcal{H} = \mathcal{N}$, we observe that $\mathop {\min }\limits_{j \in \mathcal{N}, i < T_v} \left| {{\mathcal{F}_j^{(i)}}} \right| \le q$. Select one user $p \in \cal{N}$ such that $\left| {{\mathcal{F}_p^{(\pi_{p}(i))}}} \right| \le q$ with $\pi_p()$ representing an arrangement of the set of time slots, and set ${\mathcal{H}_{p - 1}} = \mathcal{N}\backslash \left\{ {{p}} \right\}$. It is obvious that $\exists{{p - 1}} \in {\mathcal{H}_{p - 1}}$, $s.t.\left| {{F_{p - 1}^{(\pi_{p-1}(i))}}} \right| \le q$. Following such a procedure, the remained user set varies as $\mathcal{N} = {\mathcal{H}_p} \to {\mathcal{H}_{p - 1}} \to {\mathcal{H}_{p - 2}} \to  \cdots$, in which ${\mathcal{H}_{p' - 1}} = {\mathcal{H}_{p'}}\backslash \left\{ {{{p'}}} \right\}$ and $\left| {{\mathcal{F}_{p'}^{(\pi_{p'}(i))}}} \right| \le q$, $p' \in \cal{N}$. Therefore, we then obtain a user sequence $\left\{ {{1},{2}, \cdots ,{p}} \right\}$, and $p = \left| N \right|$.

When we match the time slots with the users based on the order of the above user sequence, each user ${p'}$ forms at most $q$ forbidden pairs with the former $(p'-1)$ users. Therefore, at most $q$ different time slots are matched with those users in $\mathcal{G}$, in which $j \in \mathcal{G}$ satisfying $1 \le j \le p' - 1$ and ${d_{j,p'}^{(i)}} \le {r}$. If so, user ${p'}$ is then matched with the $(q+1)$th time slot. The whole process continues until all the users are matched, i.e., a feasible solution is found for the MD-geometric SR problem given $(q+1)$ time slots. The SR problem is solvable if $T_v \ge q + 1 = \mathop {\max }\limits_{{\cal H},i} \mathop {\min }\limits_{j \in {\cal H}} \left| {\mathcal{F}_j^{(i)}} \right| +1$.

\section{Proof of the Convergence of UTSA}
The UTSA consists of two phases in which $N$ iterations are required for the first phase. Therefore, the convergence depends on the second phase, i.e., rotation matching. Given an initial matching structure $\Psi_0$ obtained from Phase 1, the structure changes as below: ${\Psi _0} \to {\Psi _1} \to {\Psi _2} \to  \cdots$. Without loss of generality, we assume that a specific rotation sequence $\mathcal{N}_s$ is considered during the structure change from $\Psi_{\lambda}$ to $\Psi_{\lambda + 1}$, i.e., ${\Psi _{\lambda  + 1}} = \Psi _{{\mathcal{N}_s},{\ell ^{opt}}}^{rot}$. Since $\ell^{opt}$ is a valid and optimal rotation matching, the relation of the total cross influence satisfies
\begin{equation}
\sum\limits_{j \in {\mathcal{N}_s}} {\sum\limits_{i \in {\Psi _{\lambda  + 1}}\left( {{j}} \right)} {Q_j^{(i)}} }  \leq                                                                                                                                                                                                                                                                                                                                                                                                                                                                                                                                                                                                                                                                                                                                                                    \sum\limits_{j \in {\mathcal{N}_s}} {\sum\limits_{i \in {\Psi _\lambda }\left( {{j}} \right)} {Q_j^{(i)}} }
\end{equation}
Therefore, the total cross influence reduces after each rotation matching. The lower bound of the cross influence exists when all the users lie outside each other's communication disks, which can be expressed by $T_v\varepsilon$.  Therefore, we can always find a rotation matching $\ell^*$ after which there are no permitted rotation matchings in the current matching structure $\Psi_{\lambda^*}$.

\section{Proof of Complexity of UTSA}
The complexity of UTSA is determined by its two phases. For the first phase, the complexity is $O\left( N \right)$ since there are $N$ iterations in total. The second phase consists of multiple iterations in each of which all valid rotation matchings are examined. We first calculate the upper bound of the number of all possible rotation matchings in each iteration, i.e., $\sum\limits_{q = 1}^{{q_{\max }}} {\mathop {qC}\nolimits_N^q } = O\left( {{N^{{q_{\max }}}}} \right)$. In practice, one iteration involves a smaller number of rotations since there exist many forbidden pairs. Therefore, given the total number of iterations $I$, the complexity can be expressed by $O\left( {I{N^{{q_{\max }}}}} \right)$.

\section{Proof of Corollary 1}
A forbidden pair $\{ j, {j'} \}$ makes a rotation matching $\ell$ invalid when $j \in {\cal{N}}^s, {j'} \notin {\cal{N}}^s$, and $\Psi \left( {{{j'}}} \right) \cap \Psi _{{\mathcal{N}^s},\ell }^{rot}\left( {{j}} \right) \ne \varnothing$, i.e., Tx users $j$ and ${j'}$ match with the same time slot after the rotation.

Given a forbidden pair $\{ j, {j'} \}$, we consider a subset of users $\mathcal{N}^s$ in which $p \in \mathcal{N}^s$ and $p \in \Psi \left( {\Psi \left( {{{j'}}} \right)} \right)$. If $\left| {{\mathcal{N}^s}} \right| = 3$, then there are $\left( N - 2 \right)$ potential invalid rotation matchings. If $\left| {{\mathcal{N}^s}} \right| = 4$, then $C\left( K-2, 2 \right)$ rotation matchings are invalid. Note that when two forbidden pairs are involved in this subset of users, at least $C\left( K-2, 2 \right)/2$ rotation matchings are invalid. Following this methods, the lower bound of the number of invalid rotation matching can be calculated as
\begin{subequations}
\begin{align}
& F\left[ {\mathop C\nolimits_{N - 2}^1  + \frac{{\mathop C\nolimits_{N - 2}^2 }}{{\left\lfloor {\left( {2 + 2} \right)/2} \right\rfloor }} + \frac{{\mathop C\nolimits_{N - 2}^3 }}{{\left\lfloor {\left( {3 + 2} \right)/2} \right\rfloor }} +  \cdots  + \frac{{\mathop C\nolimits_{N - 2}^{N - 2} }}{{\left\lfloor {\left( {N - 2 + 2} \right)/2} \right\rfloor }}} \right] \nonumber\\
= & F\left[ {\mathop C\nolimits_{N - 2}^1  + \sum\limits_{m = 1}^{\left\lfloor {\left( {N - 1} \right)/2} \right\rfloor - 1 } {\frac{2}{{2m + 2}}} \mathop C\nolimits_{N - 1}^{2m + 1}  + \frac{1}{{\left\lfloor {N/2} \right\rfloor }}\mathop C\nolimits_{N - 2}^{N - 2} } \right] \tag{22}\\
= & F\left[ {{2^{N - 1}} - 2 + \frac{1}{{\left\lfloor {N/2} \right\rfloor }}} \right]\nonumber
\end{align}
\end{subequations}

\end{appendices}

\end{document}